\documentclass[twocolumn]{aastex7}

\usepackage[utf8]{inputenc}
\usepackage{epsfig}

\newcommand{\Teff}{$T_{\rm eff}$}
\newcommand{\logg}{log($g$)}
\newcommand{\Vmic}{$\xi$}
\newcommand{\MH}{[M/H]}
\newcommand{\FeH}{[Fe/H]}

\newcommand{\Msun}{$M_{\odot}$}

\newcommand{\Ye}{$Y_{\rm e}$}
\newcommand{\Lnu}{\hat{L}_{\nu}}
\newcommand{\starname}{SMSS~0224$-$5737}

\begin{document}

\title{SMSS~J022423.27$-$573705.1: An Extremely Metal-Poor Star with the Most Pronounced Weak $r$-Process Signature}

\author[orcid=0009-0009-6151-8157,sname='Okada']{Hiroko Okada}
\email{okada@nhao.jp}
\affiliation{Nishi-Harima Astronomical Observatory, Center for Astronomy, University of Hyogo, 407-2 Nishigaichi, Sayo-cho, Sayo, Hyogo 679-5313, Japan}
\affiliation{National Astronomical Observatory of Japan, 2-21-1 Osawa, Mitaka, Tokyo 181-8588, Japan}

\author[orcid=0000-0002-8975-6829,sname='Aoki']{Wako Aoki}
\affiliation{National Astronomical Observatory of Japan, 2-21-1 Osawa, Mitaka, Tokyo 181-8588, Japan}
\affiliation{Astronomical Science Program, The Graduate University for Advanced Studies (SOKENDAI), 2-21-1 Osawa, Mitaka, Tokyo 181-8588, Japan}
\email{aoki.wako@nao.ac.jp}

\author[orcid=0000-0001-8537-3153,sname='Tominaga']{Nozomu Tominaga}
\affiliation{National Astronomical Observatory of Japan, 2-21-1 Osawa, Mitaka, Tokyo 181-8588, Japan}
\affiliation{Astronomical Science Program, The Graduate University for Advanced Studies (SOKENDAI), 2-21-1 Osawa, Mitaka, Tokyo 181-8588, Japan}
\affiliation{Department of Physics, Konan University, 8-9-1 Okamoto, Kobe, Hyogo 658-8501, Japan}
\email{nozomu.tominaga@nao.ac.jp}

\author[orcid=0000-0001-6653-8741,sname='Honda']{Satoshi Honda}
\affiliation{Nishi-Harima Astronomical Observatory, Center for Astronomy, University of Hyogo, 407-2 Nishigaichi, Sayo-cho, Sayo, Hyogo 679-5313, Japan}
\email{honda@nhao.jp}



\begin{abstract}

We present the measurement of 26 elemental abundances of SMSS~J022423.27$-$573705.1 (\starname), an extremely metal-poor (EMP) star with a weak $r$-process signature. We report the measurements of N, O, V, Zn, and Ba, and the upper limits for Mo, Ru, Pd, Ag, and Eu for the first time. \starname\ exhibits low C abundance and high N and O abundances suggesting that C is converted to N by the enhanced mixing during the evolution. 
The abundance pattern up to the Fe-peak elements is generally in good agreement with the average abundance of EMP stars, although a notable feature is the high [Zn/Fe] ratio ([Zn/Fe] $= +0.88$).
We confirm the enhancement of the first-peak neutron-capture elements (Sr, Y, and Zr) and determine a low Ba abundance [Ba/H] $= -5.25$, that is, [Ba/Fe] $= -1.45$. The extremely high ratio of [Zr/Ba] $= +2.60$ makes \starname\ the EMP star with the most pronounced weak $r$-process signature observed to date. 
The abundance pattern of the neutron-capture elements is compared with the yields from $r$-process nucleosynthesis models. The sharp decline in abundances beyond Zr disfavors neutron star merger or electron-capture supernova models, but are reproduced either by proto-neutron star wind models or by magneto-rotational supernova models. Considering the high [Zn/Fe] ratio, a magneto-rotational supernova is the most plausible origin of \starname.
This study demonstrates that the abundance measurements of both light and neutron-capture elements, even at low abundances, are crucial for unveiling the astrophysical sites of the weak $r$-process.

\end{abstract}



\section{Introduction} \label{sec:intro}

The origin of heavy elements in the universe, particularly those produced via the rapid neutron-capture process ($r$-process), remains one of the most fundamental questions in astrophysics. The $r$-process synthesizes approximately half the elements heavier than Fe, including lanthanide and actinide elements \citep{1957RvMP...29..547B}. Several astronomical objects and phenomena have been proposed as sites of $r$-process nucleosynthesis, such as the merger of binary stars, including at least one neutron star \citep{1974ApJ...192L.145L} and supernovae \citep{1994ApJ...433..229W}. 

In 2017, multi-messenger astronomy utilizing gravitational waves and electromagnetic observations first identified the electromagnetic counterpart AT~2017gfo of the neutron star merger (NSM) GW~170817 \citep[e.g.,][]{2017ApJ...848L..12A}. AT~2017gfo exhibits a rapid evolution from a blue to a red color over several days, consistent with the appearance of newly synthesized lanthanide elements with high optical opacity \citep[e.g.,][]{2017Natur.551...80K, 2017PASJ...69..102T}. Moreover, its spectra exhibit tentative signatures of Sr, La, and Ce \citep{2019Natur.574..497W, 2022ApJ...939....8D}. These probe that the NSM induces $r$-process nucleosynthesis and produces at least some of the lanthanide elements. However, this is a unique example, and it is still debated whether the NSM solely explain all $r$-process elements in the solar system and our Galaxy \citep[e.g.,][]{2018IJMPD..2742005H,2024ApJ...963..110H}. 

Traditionally, $r$-process nucleosynthesis has been studied using the abundance ratios in metal-poor stars formed from a gas enriched by one or a few nucleosynthesis event(s) in the early universe. The surface abundance of a metal-poor star preserves information on most elements synthesized in prior nucleosynthesis event(s) \citep[e.g.,][]{2005ARA&A..43..531B,2015ARA&A..53..631F}. Observations of metal-poor stars reveal the universality of the $r$-process abundance pattern, that is, the abundance patterns of $r$-process elements in most metal-poor stars agree with the solar residual $r$-process pattern \citep[][for a review]{2021RvMP...93a5002C}. Thus, most $r$-process nucleosynthesis sites are expected to synthesize both light and heavy neutron-capture elements with a robust abundance pattern, independent of the metallicity and the age of the universe. This process is called a "main" $r$-process.

On the other hand, there are several metal-poor stars that show enhancement of light $r$-process elements, such as Sr, Y, and Zr, compared to heavy $r$-process elements, such as Ba and Eu. 
An archetypal example is HD~122563, which exhibits high [Sr/Fe] and [Zr/Fe] ratios and low [Ba/Fe] and [Eu/Fe] ratios \citep{2006ApJ...643.1180H}. The abundance pattern of HD~122563 deviates from the solar residual $r$-process pattern. This suggests that a nucleosynthesis process different from the main $r$-process contributes to the chemical enrichment in the early universe \citep{2004ApJ...601..864T}. This process is called a lighter element primary process (LEPP, \citealt{2004ApJ...601..864T}), "weak" $r$-process \citep{2002PASP..114.1293T}, and/or "limited" $r$-process \citep{2018ApJ...858...92H}. Hereafter, we refer to this as the weak $r$-process.
Metal-poor stars with a weak $r$-process signature are often identified by high [Sr/Ba] and/or [Zr/Ba] ratios and show diversity in the abundance pattern of elements between the first and second peaks of $r$-process elements \citep{2017ApJ...837....8A,2021RvMP...93a5002C}.

The site of the weak $r$-process is also under debate, since the weak $r$-process pattern can be produced by truncating the main $r$-process, for example, owing to the relatively high electron fraction \citep{2015ApJ...815...82L,2015ApJ...810..109N,2021hgwa.bookE..13P}. Hence, any suggested site of the main $r$-process can reproduce a weak $r$-process pattern if the condition for the main $r$-process is not satisfied. Therefore, it is important to clarify whether the weak $r$-process occurs at an astrophysical site distinct from the main $r$-process, or under different conditions in the common site with the main $r$-process. Although detailed elemental abundance measurements would give an answer to this question, the metal-poor stars with weak $r$-process signature studied so far are relatively metal-rich with [Fe/H]\footnote{[A/B]~$= \log_{10}(N_{\rm A}/N_{\rm B})-\log_{10} (N_{\rm A}/N_{\rm B})_\odot$, where the subscript $\odot$ refers to the solar value and $N_{\rm A}$ and $N_{\rm B}$ are the abundances of elements A and B, respectively.}~$>-3$ \citep{2017ApJ...837....8A, 2018ApJ...868..110S, 2024A&A...688A.123X}. This is because extremely metal-poor (EMP) stars with [Fe/H]~$<-3$ are rare, and the measurements of low abundances of heavy $r$-process elements in stars with weak $r$-process signature are more difficult than in the $r$-process enhanced stars \citep{2020ApJS..249...30H}. 
However, stars with [Fe/H]~$>-3$ are less pristine, with a higher chance of contamination from multiple enrichment events and/or multiple nucleosynthesis processes, such as s-process. This might inhibit the firm constraint of the origin of metal-poor stars with weak $r$-process signature. On the other hand, studies have suggested that EMP stars are likely to reflect chemical signatures dominantly from only one enrichment event (\citealt{1995ApJ...451L..49A,1998ApJ...507L.135S}; see also \citealt{2013ARA&A..51..457N}).

SMSS~J022423.27$-$573705.1 (\starname) was discovered by the SkyMapper Southern Sky Survey and was subsequently confirmed to be an EMP star with [Fe/H] $\sim -4$ through high-dispersion spectroscopy \citep{2015ApJ...807..171J}. This star exhibits clear detection of the light $r$-process element Sr, with a strong enhancement of [Sr/Fe] $= +1.08$, whereas the heavy $r$-process element Ba was not detected, and an upper limit was obtained as [Ba/Fe] $< -0.91$. Consequently, \starname\ is constrained to have the highest [Sr/Ba] among the EMP stars, and is classified as an EMP star with weak $r$-process signature.

This paper reports the abundance results for \starname~based on high-resolution near-ultraviolet spectra obtained from follow-up observations. Section~\ref{sec:obs} describes the observations, and Section~\ref{sec:ana} describes the analysis method. The results are presented in Section~\ref{sec:res}, followed by discussion and summary in Sections~\ref{sec:dis} and \ref{sec:sum}, respectively.

\section{Observational data} \label{sec:obs}
Spectral data were obtained using the Very Large Telescope (VLT)/Ultraviolet and Visual Echelle Spectrograph (UVES, \citealt{2000SPIE.4008..534D}) on August 10, 2018. The reduced spectra are taken from the ESO Science portal\footnote{https://archive.eso.org/scienceportal/} (Program ID 0101.D-0677(A)). Stellar name in the database is UCAC2 6481897. The total exposure time for the spectra covering $3020$--$3880$~{\AA} and $4580$--$6680$~{\AA} is five hours, whereas that for $3730$--$5000$~{\AA} and $5650$--$9460$~{\AA} is two hours. The spectral resolutions for the blue and red ranges are approximately 40000 and 50000, respectively. 

\begin{figure*}[htbp]
\centering
\includegraphics[width=\textwidth]{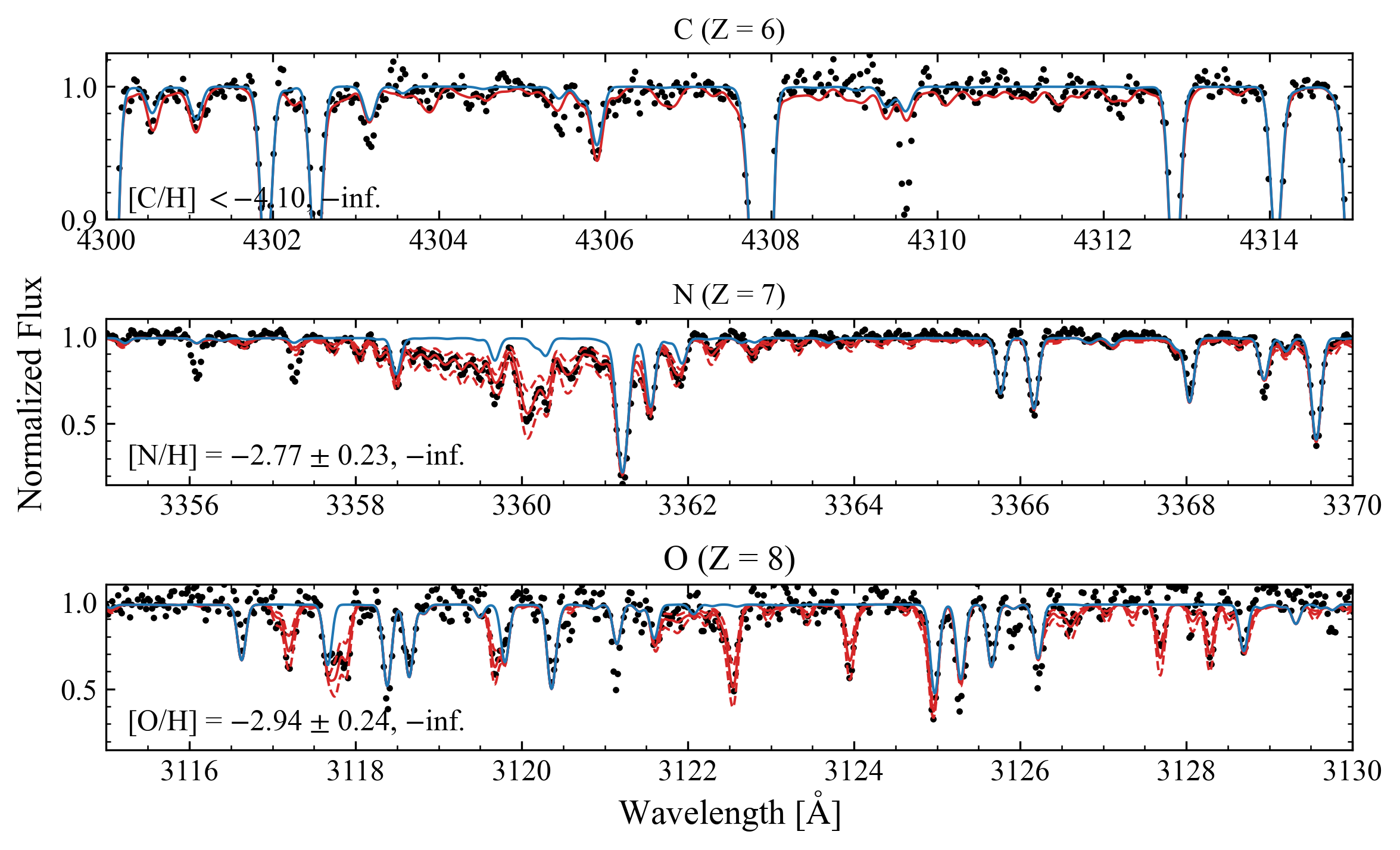}
\caption{Spectrum of \starname\ (black points) and synthetic spectra for the CH (top), NH (middle), and OH (bottom) molecular lines. The best-fit synthetic spectrum (red solid line) and synthetic spectra offset by the total errors (red dashed lines) are shown for the NH and OH lines, whereas the synthetic spectrum with an upper limit (red solid line) is shown for the CH line. The blue solid line represents the synthetic spectrum, assuming a zero abundance of each element. The adopted elemental abundances are shown in each panel.}
\label{fig:spec_CNO}
\end{figure*}

The echelle spectra are normalized and combined using the IRAF echelle package\footnote{IRAF is distributed by the National Optical Astronomy Observatories, which is operated by the Association of Universities for Research in Astronomy, Inc. under cooperative agreement with the National Science Foundation.}. The signal-to-noise ratios per pixel of the combined spectrum is 50 and 90 at 3500~{\AA} and 5000~{\AA}, respectively, by the five hour exposure, while it is 70 at 4500~{\AA} by the two hour exposure.
Examples of the reduced spectra are shown in Figures~\ref{fig:spec_CNO}(a)–(c).

\section{Analysis} \label{sec:ana}

We conduct an abundance analysis using \texttt{iSpec} \citep{2014A&A...569A.111B}.\footnote{\url{https://www.blancocuaresma.com/s/iSpec}} \texttt{iSpec} provides a unified interface for equivalent width (EW) measurements and spectral synthesis, enabling the use of various radiative transfer codes and model atmospheres. 
We employ \texttt{MOOG} \citep{2012ascl.soft02009S} for the EW-based abundance analysis and \texttt{Turbospectrum} \citep{1998A&A...330.1109A,2012ascl.soft05004P} for spectral synthesis with leveraging the strengths of both codes. While \texttt{iSpec} recommends using either \texttt{MOOG} or \texttt{WIDTH9} \citep{1993sssp.book.....K} for the EW analysis with efficiency reasons,\footnote{\url{https://www.blancocuaresma.com/s/iSpec/manual/usage/atmospheric_parameters}} 
\texttt{Turbospectrum} includes treatment of scattering
in the blue and UV domain and supports a large number of molecular lines \citep{1998A&A...330.1109A,2012ascl.soft05004P,2014A&A...569A.111B}, making it well suited for spectral synthesis including molecular lines in the wavelength range covered by this study.

We adopt the EW and spectral synthesis methods depending on the complexity of the spectral features and whether a semi-automatic measurement of EWs is available (Section~\ref{sec:ana2}). 
We compute the synthetic spectra with \texttt{Turbospectrum} adopting the abundances derived from the EW methods with \texttt{MOOG} and compare the shapes of lines in the observed spectra, which are used for the abundance measurement with the EW methods, with those of the synthetic spectra.
The line shapes are mostly consistent, confirming that both analysis methods yield consistent abundance results.\footnote{The differences on abundance estimates, which are caused by the selection of adopting codes, are investigated with {\it Gaia} benchmark stars in \cite{2019MNRAS.486.2075B}. The absolute values of median differences are mostly $\lesssim0.2$~dex.}
This agreement demonstrates that our derived abundances are robust against the choice of analysis methods and radiative transfer codes.

All the analyses are conducted under the assumption of local thermodynamic equilibrium (LTE), using a one-dimensional hydrostatic model atmosphere \texttt{MARCS} \citep{2008A&A...486..951G}. 
We adopt the fine interpolated version of the spherical model atmosphere grids in \texttt{iSpec} \citep{2014A&A...569A.111B}, which is computed from the original \texttt{MARCS} model atmosphere grids with typical steps on effective temperature of $\sim250$~K, surface gravity of $\sim0.5$~dex, and metallicity of $\sim1$~dex around the stellar parameters of \starname\ (Section~\ref{sec:ana1}). The differences on abundance estimates, which are introduced by the interpolation, are typically as small as $\sim0.05$~dex \citep{2014A&A...569A.111B}. 
We adopt solar abundances from \citet{2009ARA&A..47..481A}.

The atomic line data are mainly obtained from the Vienna Atomic Line Database (VALD; \citealp{2011BaltA..20..503K}). For the molecular transitions, we adopt the CH line lists from \citet{MPV},\footnote{A modified version on Mar 16, 2017 is adopted.} NH from \citet{2018JQSRT.217...29F}, and OH from \citet{KOH}. We use the data for \ion{Ba}{2} and \ion{Eu}{2} lines from \citet{MW,Wc,CB,ESTM} and \citet{LWHS}, respectively. In this study, solar isotopic ratios are adopted for all elements.

\subsection{Stellar Atmospheric Parameters} \label{sec:ana1}

The stellar atmospheric parameters of \starname\ are determined as follows: A photogeometric distance of 5,380~pc is adopted from the Gaia EDR3 \citep{2016A&A...595A...1G,2021A&A...649A...1G,2021AJ....161..147B}. Using this distance, in combination with the Galactic dust reddening map of \citet{2011ApJ...737..103S} and assuming a $\mathrm{sech}^2$ vertical dust distribution with a scale height of 125pc \citep{2006A&A...453..635M}, we derive a color excess of $E(B-V) = 0.037$.\footnote{Since \starname\ lies 4343~pc from the Galactic plane, the reddening is independent of the assumed dust distribution.}

The intrinsic $V-K_s$ color is calculated using $V = 13.469$ (APASS DR10)\footnote{\url{https://www.aavso.org/apass}} and $K_s = 11.332$ (2MASS; \citealp{2006AJ....131.1163S}), corrected for interstellar extinction \citep{1992ApJ...395..130P}. By applying the infrared flux method \citep{1999A&AS..140..261A} and assuming [Fe/H] = $-3$, we derive an effective temperature of $T_{\mathrm{eff}} = 5039$~K. The bolometric correction is performed using the same method. Assuming a stellar mass of 0.8\Msun, the surface gravity is determined to be \logg\ = 1.96 using the distance and \Teff. \starname\ is located at the red giant branch in the \Teff--\logg\ diagram.

The model atmosphere is interpolated for \Teff, \logg, and \MH\ from the \texttt{MARCS} models. We determine the microturbulent velocity $\xi$ using \texttt{MOOG} by minimizing the trend between the Fe abundances and reduced equivalent widths from the \ion{Fe}{1} lines. A total of 96 \ion{Fe}{1} and seven \ion{Fe}{2} lines are analyzed. 
Starting from an initial assumption of \MH\ = 0, we derive a preliminary estimate of $\xi$ and \MH\ and redetermine the parameters with the preliminary estimate. Here, we normalize the observed spectra with the continuum flux estimated by taking the median flux of "continuum pixels," where no strong absorption lines are present in synthetic spectra, over 5~\AA\ windows and interpolating these values using a piecewise cubic Hermite interpolating polynomial (PCHIP) one-dimensional monotonic cubic interpolation. The final values are $\xi = 1.90$~km~s$^{-1}$ and \MH\ = $-3.80$.

Compared with \citet{2015ApJ...807..171J}, which derives \Teff\ $=4846$~K, \logg\ $= 1.6$, and [Fe/H] = $-3.97$ from the spectroscopic analysis, our metallicity is higher by 0.17dex. This difference results from a higher \Teff\ by 193~K and a higher \logg\ by 0.33~dex in our analysis.

To validate our temperature estimate, we apply several photometric color–$T_{\mathrm{eff}}$ relations from \citet{1999A&AS..140..261A} and \citet{2021A&A...653A..90M}. The $B-V$, $J-H$, $J-K_s$, $BP-RP$, $BP-G$, $G-RP$, $BP-K_s$, $RP-K_s$, and $G-K_s$ colors of \starname\ result in \Teff~$=5264$, $5275$, $5384$, $5170$, $5175$, $5166$, $5188$, $5205$, and $5191$~K, respectively. All these result in a higher \Teff\ than that in \cite{2015ApJ...807..171J}. We adopt \Teff$=5039$~K, \logg\ $= 1.96$~dex, and [M/H] $=-3.80$ in this study because the $V-K_s$ color gives the smallest dispersion among these colors and because the \logg\ estimates using the distance derived from the Gaia parallax are more reliable than the standard spectroscopic techniques \citep[e.g.,][]{2022ApJ...931..147L}. Systematic errors caused by errors in the stellar parameters are described in Section~\ref{sec:ana3}.

\subsection{Abundance Analysis} \label{sec:ana2}

\subsubsection{Measurements with Equivalent Widths}

The abundances of elements with relatively isolated atomic lines are derived using a standard analysis of EWs. The EWs are measured semiautomatically with \texttt{iSpec} by fitting Gaussian or Lorentzian profiles to absorption lines in the spectra normalized to the continuum flux (Section~\ref{sec:ana1}).

All selected absorption lines are visually inspected to confirm their reliability. Lines blended with the absorption features of other elements, spurious features caused by artifacts, telluric contamination, or uncertain continuum placement, are excluded from the analysis.

\subsubsection{Measurements with Synthetic Spectra}

The abundances are determined by comparing the observed spectra with the synthetic spectra for molecules exhibiting numerous absorption lines, as well as for atomic lines for which EWs cannot be semi-automatically measured. The best-fit abundances are obtained by minimizing the $\chi^2$ value derived from the flux residuals between the observed and synthetic spectra over the pixels containing the absorption lines.

For molecular features, we synthesize spectra with the best-fit stellar atmospheric parameters and varying abundances. Adopting the normalized synthetic spectrum, we identify the continuum pixels and "molecular-line pixels," to which molecular lines dominantly contribute. The observed spectra are normalized using the continuum flux, which is estimated using the continuum pixels. $\chi^2$ values are evaluated using the flux residuals between the normalized observed and synthetic spectra over the molecular-line pixels. The best-fit abundance is derived to minimize $\chi^2$.

\subsubsection{Upper Limits}

For elements without significant absorption features, we estimate the $3\sigma$ upper limit from the expected uncertainty of the equivalent widths using the equation in \citet{1988IAUS..132..345C}. We adopt synthetic spectra to convert the expected uncertainty into upper limits of abundance.

\subsection{Error Estimation} \label{sec:ana3}

The random error in the measurement with EWs is estimated to be $\sigma N^{-1/2}$, where $\sigma$ is the standard deviation of the derived abundances from individual lines, and $N$ is the number of lines used. The $\sigma$ of \ion{Fe}{1} lines is adopted to estimate the random errors for ions with four or fewer available lines (\ion{Al}{1}, \ion{Si}{1}, \ion{Ca}{1}, \ion{Mn}{1}, \ion{Zn}{1}, and \ion{Sr}{2}) and \ion{Fe}{2} with $\sigma$ smaller than that of \ion{Fe}{1} lines.

The random error of an atomic line in the measurement with synthetic spectra is estimated from the error in the continuum placement, that is, the $S/N$ ratio of the continuum. The associated error depends on the $S/N$ ratio of the spectrum and the line strength. Because there are numerous absorption lines of molecules, we adopt $\sigma$ of the \ion{Fe}{1} lines as the error of the abundance of molecules.
The observed spectrum is located between the synthetic spectra derived with the resultant errors (Figures~\ref{fig:spec_CNO})

The errors resulting from the uncertainties in the stellar parameters are estimated by varying \Teff\ by $\pm 100$~K, \logg\ and \FeH\ by $\pm 0.3$~dex, and \Vmic\ by $\pm 0.3$~km~s$^{-1}$. These errors are listed in Table~\ref{tab:err}. The total error is obtained by adding the random error in the quadrature and the errors due to the uncertainties of the stellar parameters.

\begin{deluxetable*}{crrrrrrrrr}
\tabletypesize{\scriptsize}
\tablecaption{Error estimates for \starname. \label{tab:err}}
\tablewidth{0pt}
\tablehead{
\colhead{} &
\colhead{} &
\multicolumn{2}{c}{$\Delta T_{\rm eff}$~[K]} &
\multicolumn{2}{c}{$\Delta \log g$~[dex]} &
\multicolumn{2}{c}{$\Delta \rm [Fe/H]$~[dex]} &
\multicolumn{2}{c}{$\Delta \xi$~[km~s$^{-1}$]} \\
\colhead{Element} &
\colhead{$\sigma_{\rm ran}$} &
\colhead{$-100$} & \colhead{$+100$} &
\colhead{$-0.3$} & \colhead{$+0.3$} &
\colhead{$-0.3$} & \colhead{$+0.3$} &
\colhead{$-0.3$} & \colhead{$+0.3$}
}
\startdata
NH&$+0.09$&$-0.13$&$+0.23$&$+0.10$&$-0.11$&$+0.05$&$+0.04$&$-0.01$&$+0.00$\\
OH&$+0.09$&$-0.16$&$+0.23$&$+0.12$&$-0.10$&$+0.03$&$+0.04$&$+0.02$&$+0.01$\\
\ion{Na}{1}&$+0.07$&$-0.10$&$+0.09$&$+0.01$&$-0.02$&$+0.00$&$+0.01$&$+0.04$&$-0.03$\\
\ion{Mg}{1}&$+0.06$&$-0.06$&$+0.06$&$+0.02$&$+0.00$&$-0.00$&$+0.00$&$+0.08$&$-0.02$\\
\ion{Al}{1}&$+0.09$&$-0.10$&$+0.09$&$+0.02$&$-0.03$&$+0.00$&$+0.00$&$+0.08$&$-0.06$\\
\ion{Si}{1}&$+0.09$&$-0.10$&$+0.10$&$+0.06$&$-0.06$&$-0.01$&$+0.00$&$+0.18$&$-0.18$\\
\ion{Ca}{1}&$+0.04$&$-0.07$&$+0.06$&$+0.01$&$-0.02$&$+0.00$&$+0.00$&$+0.02$&$-0.01$\\
\ion{Sc}{2}&$+0.05$&$-0.07$&$+0.07$&$-0.06$&$+0.08$&$-0.00$&$+0.00$&$+0.05$&$-0.03$\\
\ion{Ti}{1}&$+0.04$&$-0.13$&$+0.12$&$+0.02$&$-0.04$&$+0.00$&$+0.00$&$+0.01$&$-0.01$\\
\ion{Ti}{2}&$+0.03$&$-0.05$&$+0.07$&$-0.06$&$+0.08$&$-0.00$&$+0.01$&$+0.04$&$-0.09$\\
\ion{V}{2}&$+0.10$&$-0.05$&$+0.05$&$-0.06$&$+0.07$&$-0.00$&$+0.00$&$+0.03$&$-0.02$\\
\ion{Cr}{1}&$+0.19$&$-0.11$&$+0.10$&$+0.00$&$-0.02$&$+0.00$&$+0.00$&$+0.01$&$+0.04$\\
\ion{Cr}{2}&$+0.05$&$-0.02$&$+0.03$&$-0.06$&$+0.08$&$-0.00$&$+0.00$&$+0.04$&$-0.01$\\
\ion{Mn}{2}&$+0.06$&$-0.04$&$+0.04$&$-0.06$&$+0.08$&$-0.00$&$+0.00$&$+0.02$&$-0.02$\\
\ion{Mn}{1}&$+0.05$&$-0.14$&$+0.13$&$+0.01$&$-0.03$&$+0.00$&$+0.00$&$+0.01$&$-0.01$\\
\ion{Fe}{1}&$+0.01$&$-0.10$&$+0.09$&$+0.02$&$-0.03$&$+0.00$&$+0.00$&$+0.07$&$-0.04$\\
\ion{Fe}{2}&$+0.03$&$-0.02$&$+0.02$&$-0.08$&$+0.11$&$-0.01$&$+0.00$&$+0.05$&$-0.02$\\
\ion{Co}{1}&$+0.02$&$-0.14$&$+0.13$&$+0.02$&$-0.04$&$+0.00$&$+0.00$&$+0.03$&$-0.02$\\
\ion{Ni}{1}&$+0.02$&$-0.13$&$+0.13$&$+0.04$&$-0.04$&$-0.00$&$+0.00$&$+0.14$&$-0.10$\\
\ion{Zn}{1}&$+0.19$&$-0.04$&$+0.05$&$-0.02$&$+0.03$&$-0.00$&$+0.00$&$+0.00$&$-0.00$\\
\ion{Sr}{2}&$+0.06$&$-0.11$&$+0.10$&$-0.01$&$+0.00$&$-0.01$&$+0.00$&$+0.24$&$-0.27$\\
\ion{Y}{2}&$+0.02$&$-0.06$&$+0.08$&$-0.07$&$+0.11$&$-0.00$&$+0.00$&$+0.04$&$-0.00$\\
\ion{Zr}{2}&$+0.02$&$-0.07$&$+0.07$&$-0.07$&$+0.07$&$-0.00$&$+0.00$&$+0.04$&$-0.04$\\
\ion{Ba}{2}&$+0.18$&$-0.09$&$+0.10$&$-0.08$&$+0.11$&$-0.01$&$+0.01$&$+0.00$&$+0.00$\\
\enddata
\end{deluxetable*}

\section{Results} \label{sec:res}

\subsection{Abundance Measurement} \label{sec:res:abn}

The resulting elemental abundances and total errors are listed in Table~\ref{tab:abun}.

\begin{deluxetable*}{clcrrrcc}
\tabletypesize{\scriptsize}
\tablecaption{Elemental abundances of \starname. \label{tab:abun}}
\tablewidth{0pt}
\tablehead{
\colhead{Z} &
\colhead{Element} &
\colhead{Sun $\log \epsilon$(X)} &
\colhead{$\log \epsilon$(X)} &
\colhead{[X/H]} &
\colhead{[X/Fe]} &
\colhead{Error} &
\colhead{N}
}
\startdata
6 & CH             & $+8.43$ & $\textless~4.33$ & $\textless -4.10$ & $\textless -0.31$ & \textemdash & \textit{Synthesis} \\
$7$&NH&$+7.83$&$+5.06$&$-2.77$&$+1.03$&$0.23$&\textit{Synthesis} \\
$8$&OH&$+8.69$&$+5.75$&$-2.94$&$+0.86$&$0.24$&\textit{Synthesis} \\
$11$&\ion{Na}{1}&$+6.24$&$+2.43$&$-3.81$&$-0.01$&$0.12$&$2$ (\textit{Synthesis}) \\
$12$&\ion{Mg}{1}&$+7.60$&$+4.43$&$-3.17$&$+0.63$&$0.10$&$7$ \\
$13$&\ion{Al}{1}&$+6.45$&$+2.33$&$-4.12$&$-0.32$&$0.15$&$1$ \\
$14$&\ion{Si}{1}&$+7.51$&$+4.25$&$-3.26$&$+0.53$&$0.23$&$1$ \\
$20$&\ion{Ca}{1}&$+6.34$&$+2.87$&$-3.47$&$+0.33$&$0.08$&$4$ \\
$21$&\ion{Sc}{2}&$+3.15$&$-0.34$&$-3.49$&$+0.31$&$0.12$&$19$ \\
$22$&\ion{Ti}{1}&$+4.95$&$+1.50$&$-3.45$&$+0.35$&$0.14$&$9$ \\
{}&\ion{Ti}{2}&$+4.95$&$+1.60$&$-3.35$&$+0.44$&$0.12$&$75$ \\
$23$&\ion{V}{2}&$+3.93$&$+0.43$&$-3.50$&$+0.30$&$0.13$&$8$ \\
$24$&\ion{Cr}{1}&$+5.64$&$+1.55$&$-4.09$&$-0.30$&$0.22$&$11$ \\
{}&\ion{Cr}{2}&$+5.64$&$+1.95$&$-3.69$&$+0.11$&$0.09$&$9$ \\
$25$&\ion{Mn}{1}&$+5.43$&$+0.73$&$-4.70$&$-0.90$&$0.14$&$3$ \\
{}&\ion{Mn}{2}&$+5.43$&$+1.33$&$-4.10$&$-0.30$&$0.10$&$5$ \\
$26$&\ion{Fe}{1}&$+7.50$&$+3.70$&$-3.80$&$+0.00$&$0.11$&$96$ \\
{}&\ion{Fe}{2}&$+7.50$&$+3.82$&$-3.68$&$+0.12$&$0.11$&$7$ \\
$27$&\ion{Co}{1}&$+4.99$&$+1.67$&$-3.32$&$+0.48$&$0.15$&$38$ \\
$28$&\ion{Ni}{1}&$+6.22$&$+2.62$&$-3.60$&$+0.20$&$0.18$&$45$ \\
$30$&\ion{Zn}{1}&$+4.56$&$+1.60$&$-2.96$&$+0.88$&$0.20$&$2$ (\textit{EW, Synthesis}) \\
$38$&\ion{Sr}{2}&$+2.87$&$+0.14$&$-2.73$&$+1.06$&$0.29$&$2$ \\
$39$&\ion{Y}{2}&$+2.21$&$-0.91$&$-3.12$&$+0.67$&$0.11$&$15$ \\
$40$&\ion{Zr}{2}&$+2.58$&$-0.07$&$-2.65$&$+1.15$&$0.11$&$26$ \\
$42$ & \ion{Mo}{1}   & $+1.88$ & $\textless -0.42$ & $\textless -2.30$ & $\textless +1.49$ & \textemdash & \textit{Upper Limit} \\
$44$ & \ion{Ru}{1}   & $+1.75$ & $\textless -0.20$ & $\textless -1.95$ & $\textless +1.84$ & \textemdash & \textit{Upper Limit} \\
$46$ & \ion{Pd}{1}   & $+1.65$ & $\textless -0.13$ & $\textless -1.70$ & $\textless +2.09$ & \textemdash & \textit{Upper Limit} \\
$47$ & \ion{Ag}{1}   & $+1.69$ & $\textless -0.96$ & $\textless -1.90$ & $\textless +1.89$ & \textemdash & \textit{Upper Limit} \\
$56$&\ion{Ba}{2}&$+2.18$&$-3.07$&$-5.25$&$-1.45$&$0.23$&$1$ (\textit{Synthesis}) \\
$63$ & \ion{Eu}{2}   & $+0.52$ & $\textless -2.53$ & $\textless -3.05$ & $\textless +0.74$ & \textemdash & \textit{Upper Limit}
\enddata
\end{deluxetable*}

The C, N, and O abundances are derived from the molecular bands of CH lines at 4310~\AA, NH lines at 3360~\AA, and OH lines at 3100~\AA, respectively. Figures~\ref{fig:spec_CNO}(a)–(c) show the normalized observed spectra, where the transitions of the CH, NH, and OH molecular lines exist. The NH and OH lines are well reproduced with [N/H] $=-2.77$, that is, [N/Fe] $=+1.03$, and [O/H] $=-2.94$, that is, [O/Fe] $=+0.86$, whereas the absence of CH lines is consistent with the spectrum and the upper limit of [C/H] $<-4.10$, that is, [C/Fe] $<-0.31$, is derived. Here, we derive the upper limit of [C/H] by requiring that the peak depths of the CH lines be shallower than three times the standard deviation of the continuum with a signal-to-noise ratio of $\sim145$. Although the C abundance is measured as [C/Fe]~$=+0.07$ in \cite{2015ApJ...807..171J}, we adopt the upper limit because of the higher $S/N$ ratios in our analysis.

The abundances of Mg, Al, Si, Ca, Sc, Ti, V, Cr, Mn, Fe, Co, Ni, Sr, Y, and Zr are determined using the EW methods. Most elements are measured using multiple lines, yielding robust statistics. However, the number of available lines is small for Si, Al, and Sr. We use the lines at 3905~\AA\ for Si, 3961~\AA\ for Al, and 4077 and 4215~\AA\ for Sr. 

We separately estimate the abundances of Ti, Cr, Mn, and Fe in the neutral and singly ionized stages. There are discrepancies between the abundances estimated for the two states of Cr and Mn. This stems from non-LTE (NLTE) effects on the \ion{Cr}{1} and \ion{Mn}{1} lines, which are prominent at low metallicity \citep{2008A&A...492..823B,2010A&A...522A...9B}. Adopting the NLTE corrections of $+0.3$--$+0.5$~dex for \ion{Cr}{1} and $+0.5$--$+0.7$~dex for \ion{Mn}{1} in metal-poor stars, the abundances measured with the \ion{Cr}{1} and \ion{Mn}{1} lines are consistent with those measured with the \ion{Cr}{2} and \ion{Mn}{2} lines, respectively. Therefore, we adopt measurements with \ion{Cr}{2} and \ion{Mn}{2} lines as the abundances of Cr and Mn, respectively.

Hyperfine splitting (HFS) affects certain odd-Z species such as Sc, V, Mn, and Co \citep[see][for a brief review]{2018Atoms...6...48S}. The line list including the effects of HFS on Sc \citep{LD}, V \citep{WLDSC}, Mn \citep{DLSSC,MFW}, and Co \citep{FMW} are adopted in the analysis. 

Notably, the Sr lines may be too strong for certain abundance measurements. Hence, we adopt Zr instead of Sr as an indicator of excess light neutron-capture elements. 

The abundances of Na and Ba are measured using the synthetic spectra method. 
We adopt \ion{Na}{1} lines at 5890 and 5896~\AA. Because these lines are strong, the errors due to continuum placement are as low as 0.03~dex and the difference between the measurements from these lines is 0.08~dex. The resulting random error is 0.07~dex. The resultant abundance of Na is [Na/H] = $-3.81$, i.e., [Na/Fe] = $-0.01$.

\begin{figure}[t]
\centering
\includegraphics[width=\columnwidth]{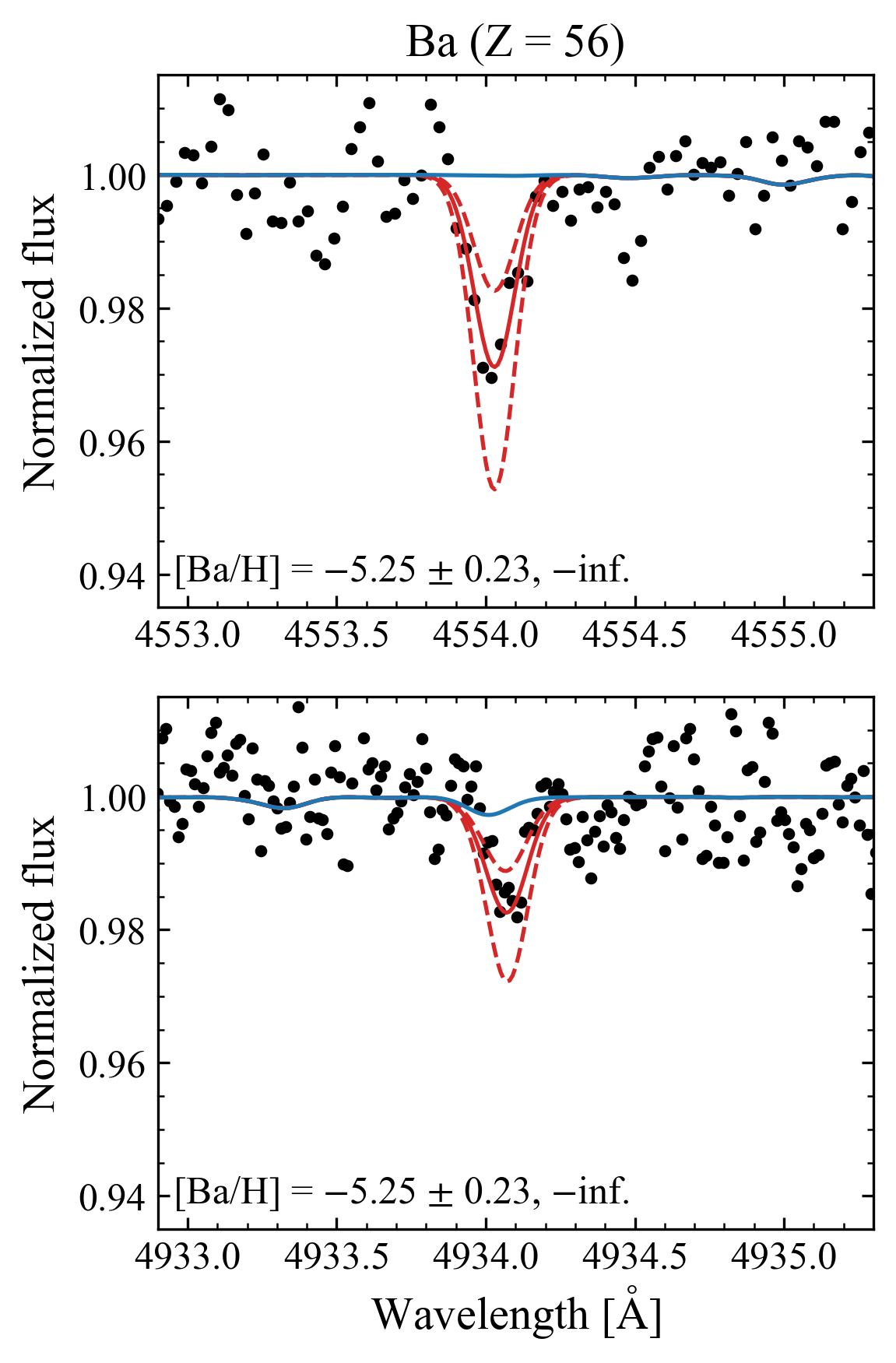}
\caption{Same as Figure~\ref{fig:spec_CNO}, but for the spectrum of \starname\ and the synthetic spectra for the \ion{Ba}{2} lines at 4554~\AA\ and 4934 \AA. The best-fit abundance is [Ba/H] = $-5.25 \pm 0.23$.}
\label{fig:spec_Ba}
\end{figure}

\begin{figure}[t]
\centering
\includegraphics[width=\columnwidth]{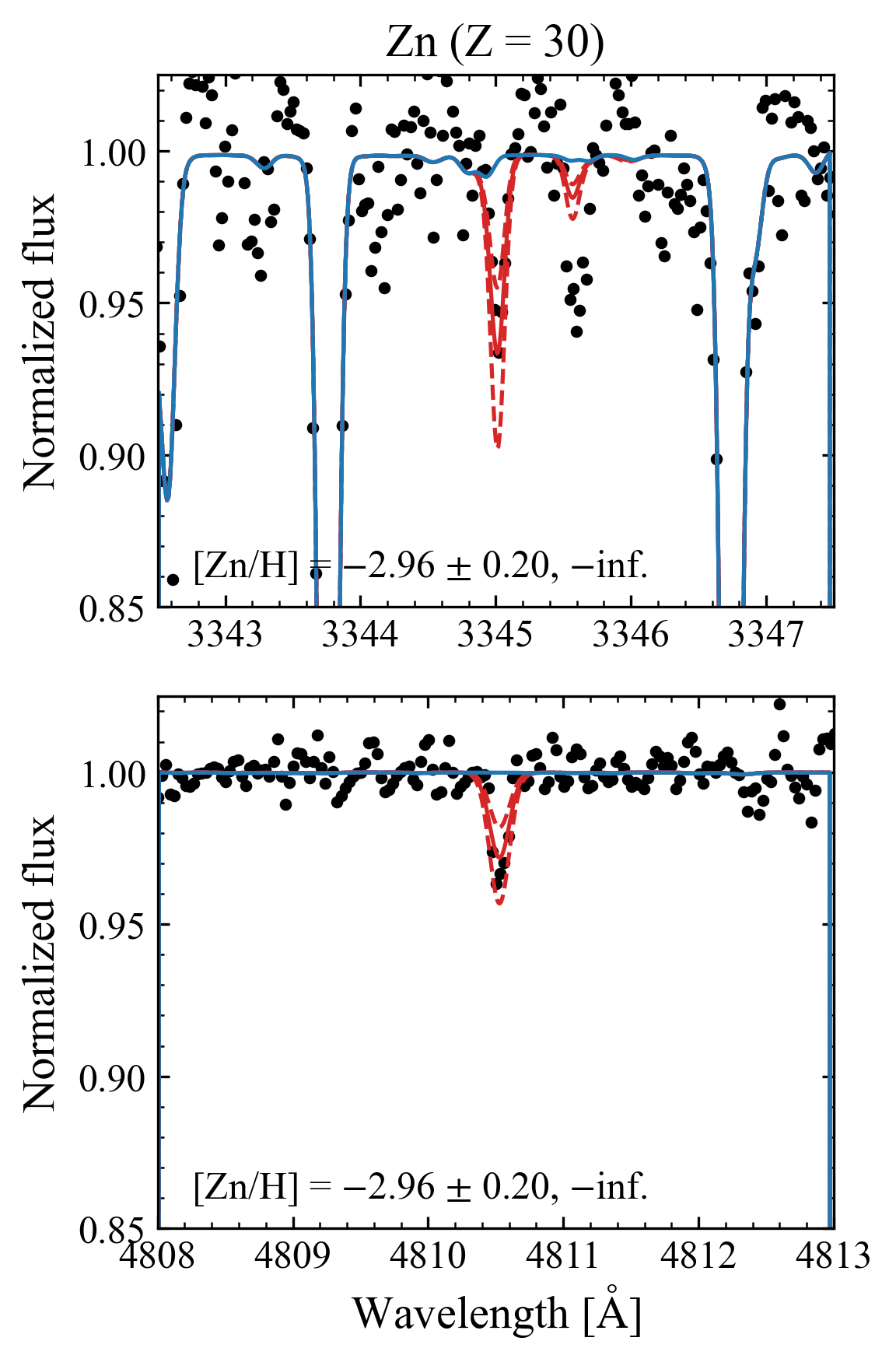}
\caption{Same as Figure~\ref{fig:spec_CNO}, but for the spectrum of \starname\ and the synthetic spectra for the \ion{Zn}{1} lines at 3345~\AA\ and 4810~\AA. The best-fit abundance is [Zn/H] = $-2.96 \pm 0.20$.}
\label{fig:spec_Zn}
\end{figure}

To measure the Ba abundance, we use the strong \ion{Ba}{2} resonance line at 4554~\AA, which is the only accessible transition for Ba in most EMP stars. 
This line is detected at a significance of $4 \sigma$ because of the high signal-to-noise ratio of continuum of $\sim 100$ (the upper panel of Figure~\ref{fig:spec_Ba}). 
We adopt the line transition from \citet{MW} without the effect of the HFS. The synthetic spectrum fits well the spectral line. The best-fit abundance is [Ba/H] = $-5.25$, that is, [Ba/Fe] = $-1.45$. The uncertainty due to the continuum placement, 0.16~dex, is smaller than errors due to noise at this wavelength. To estimate the HFS effect, we calculate the synthetic spectra using the line list including HFS effects \citep{1998AJ....115.1640M} assuming the isotope ratios of the r-process component in solar-system material. The difference between the abundance derived with and without the HFS effects is  smaller than $0.05$~dex, less than the uncertainty due to other causes.
We also confirm that the derived Ba abundance is consistent with the \ion{Ba}{2} resonance line at 4934~\AA\ (the lower panel of Figure~\ref{fig:spec_Ba}). 

We adopt the EW and spectral synthesis methods to measure the Zn abundance, using two \ion{Zn}{1} lines at 3345 and 4810~\AA. Because the line at 3345~\AA\ in the UV region is detected and its EW is semi-automatically measured, the abundance, [Zn/H]~$=-2.98$, is determined with the EW method. The line at 4810~\AA\ is weaker than at 3345~\AA\ and not semi-automatically detected, but its high S/N ratio allows the measurement of the abundance, [Zn/H]~$=-2.93$, with the spectral synthesis method. The uncertainty of the abundance measurement with the spectral synthesis due to the continuum placement is $0.18$~dex and dominates the random error of the Zn abundance.
The resultant abundance of Zn is [Zn/H] = $-2.96$, i.e., [Zn/Fe] = $+0.88$.
The agreement of the abundances from the two lines confirms the robustness of the measurement of the Zn abundance with the two \ion{Zn}{1} lines (Figure~\ref{fig:spec_Zn}). Although there is a weak feature of Zn at 3345.6~\AA, the line may be overlapped with an unidentified line that is also visible in a spectrum of other metal-poor giant stars.

\begin{figure*}[htbp]
\centering
\includegraphics[width=\textwidth]{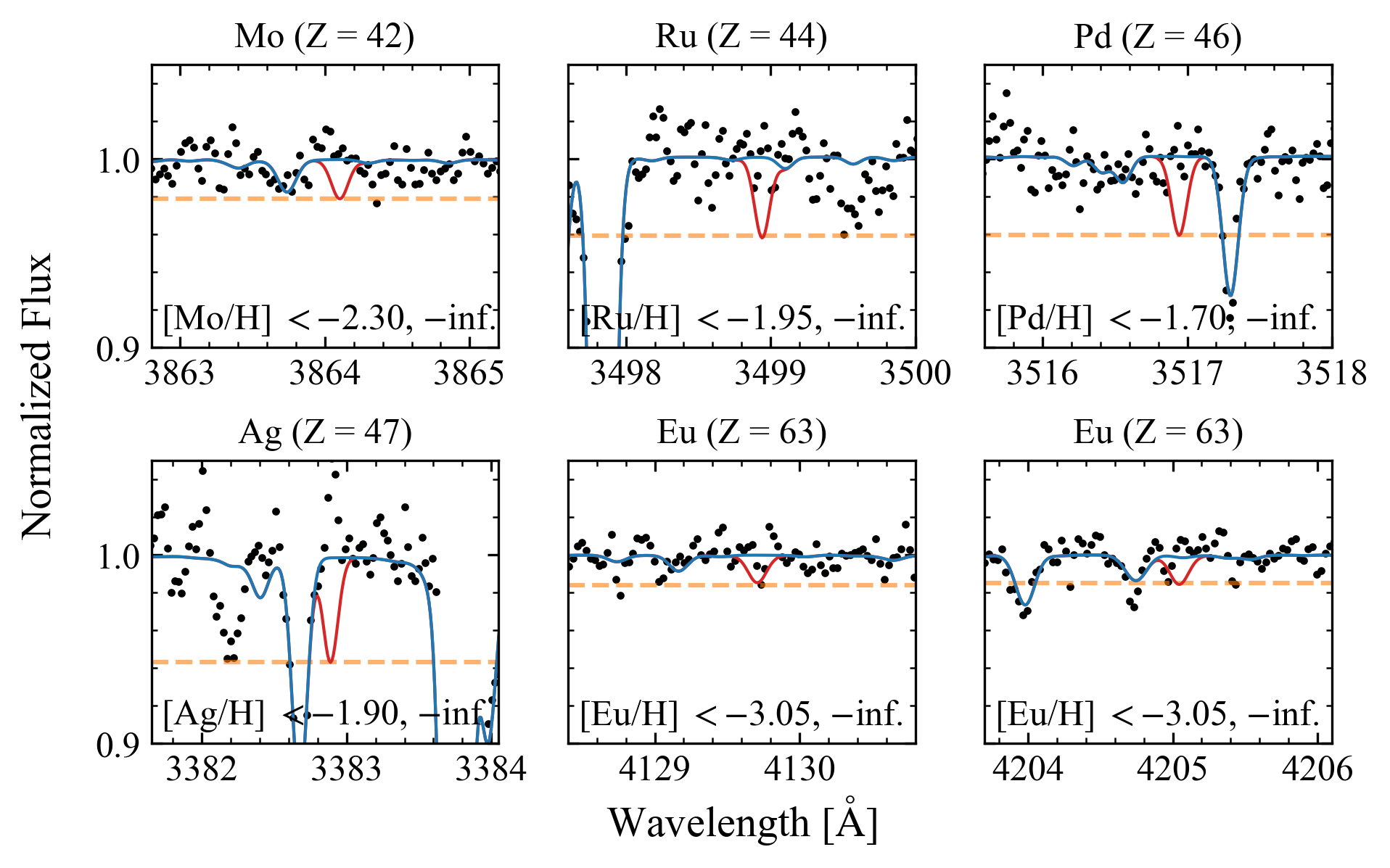}
\caption{Spectrum of \starname\ (black points) and synthetic spectra at the strongest transitions of the neutron-capture elements (Mo, Ru, Pd, Ag, and Eu). The blue solid lines show the synthetic spectra assuming a zero abundance for each element, whereas the red solid lines represent the synthetic spectra with an upper limit. The orange dashed horizontal lines indicate the $3 \sigma$ detection thresholds estimated from the local signal-to-noise ratio (S/N). No significant lines are detected, and only the upper limits are derived.}
\label{fig:spec_UL}
\end{figure*}

No significant absorption is detected for heavier light neutron-capture elements such as Mo, Ru, Pd, Ag, and Eu (Figure~\ref{fig:spec_UL}). Using the synthetic spectra and local signal-to-noise ratio, we derive $3\sigma$ upper limits for their strongest transitions: \ion{Mo}{1}~3864~\AA, \ion{Ru}{1}~3499~\AA, \ion{Pd}{1}~3404~\AA, \ion{Ag}{1}~3280~\AA, and \ion{Eu}{2}~4129 and 4205~\AA. The upper limits are [Mo/H] $< -2.30$, [Ru/H] $< -1.95$, [Pd/H] $< -1.70$, [Ag/H] $< -1.90$, and [Eu/H] $<-3.05$.

\subsection{Characteristics of Elemental Abundances} \label{sec:res:char}

In this subsection, the elemental abundances of \starname\ are compared with those of other EMP stars and its characteristics are summarized. 

\begin{figure*}[t]
\centering
\includegraphics[width=\textwidth]{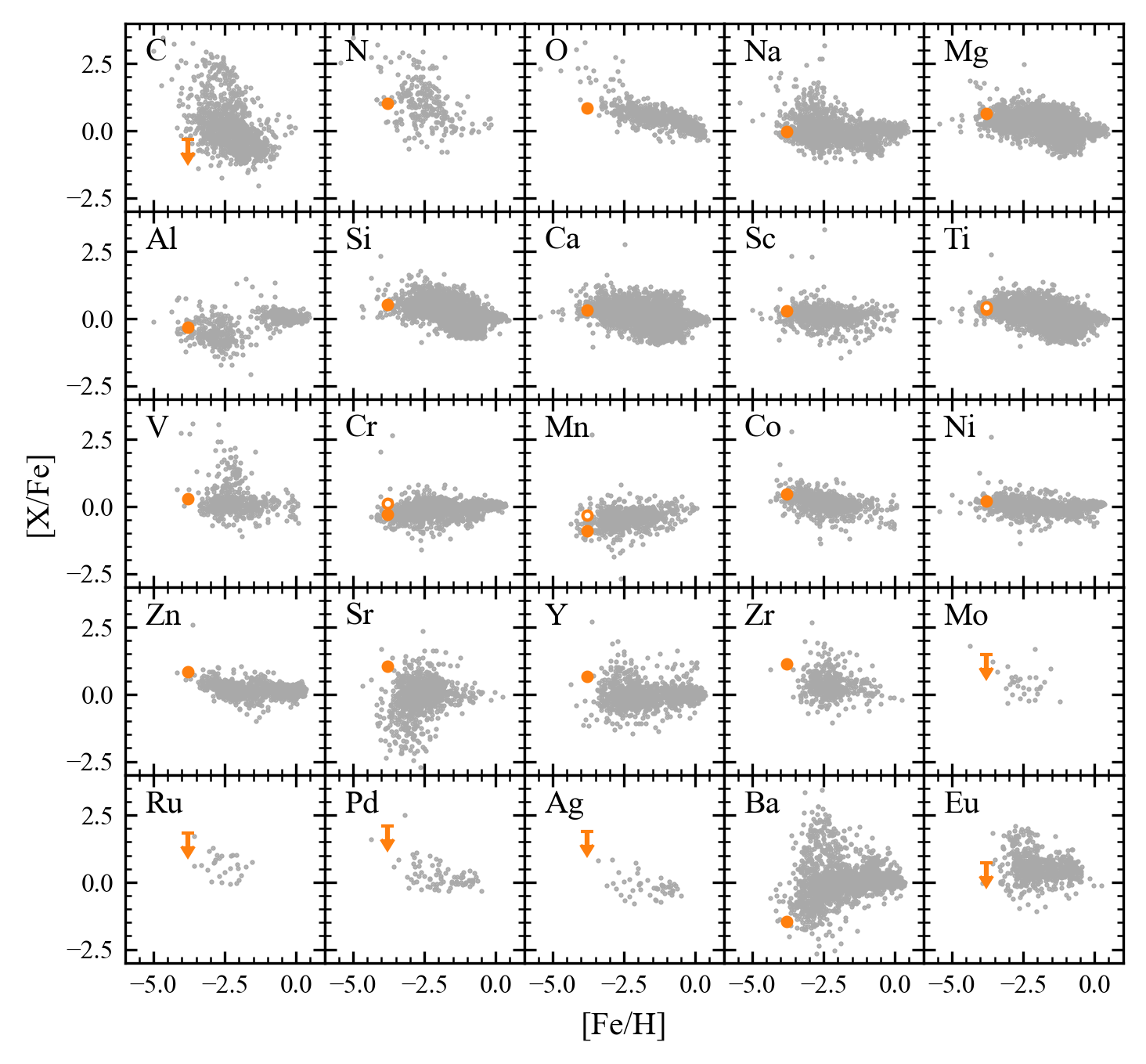}
\caption{Elemental abundance ratios [X/Fe] as a function of metallicity of elements from C to Eu. The grey points represent the abundance ratios of stars compiled by the SAGA database \citep{2008PASJ...60.1159S}. The elemental abundance ratios and upper limits of \starname\ are indicated by orange symbols. When the abundance estimates from the two ionization levels are available, the abundance ratios derived from singly-ionized ions are shown as open circles.}
\label{fig:abun_all}
\end{figure*}

Figure~\ref{fig:abun_all} shows the abundance ratios [X/Fe] as a function of metallicity [Fe/H] compared to stars collected in the SAGA database \citep{2008PASJ...60.1159S}. 
The abundance ratios [X/Fe] up to the Fe-peak elements in \starname\ are consistent with those of the observed EMP stars. The high abundances of Co and Zn are also consistent with the trend among the EMP stars \citep[e.g.,][]{2019ApJ...876...97E}. These results indicate that the abundance pattern of \starname\ is typical of EMP stars with [Fe/H] $\sim -4$.
However, the abundances of the neutron-capture elements in \starname\ are located in a distinctive region relative to those of EMP stars. The abundance ratios of light neutron-capture elements [(Sr, Y, Zr)/Fe] are significantly enhanced and located at the upper ends of the distribution of EMP stars. The Ba abundance in \starname\ is located at the lower end of the distribution of EMP stars, and the Eu abundance, although given as an upper limit, is lower than that of half of the EMP stars.

\begin{figure}[t]
\centering
\includegraphics[width=\columnwidth]{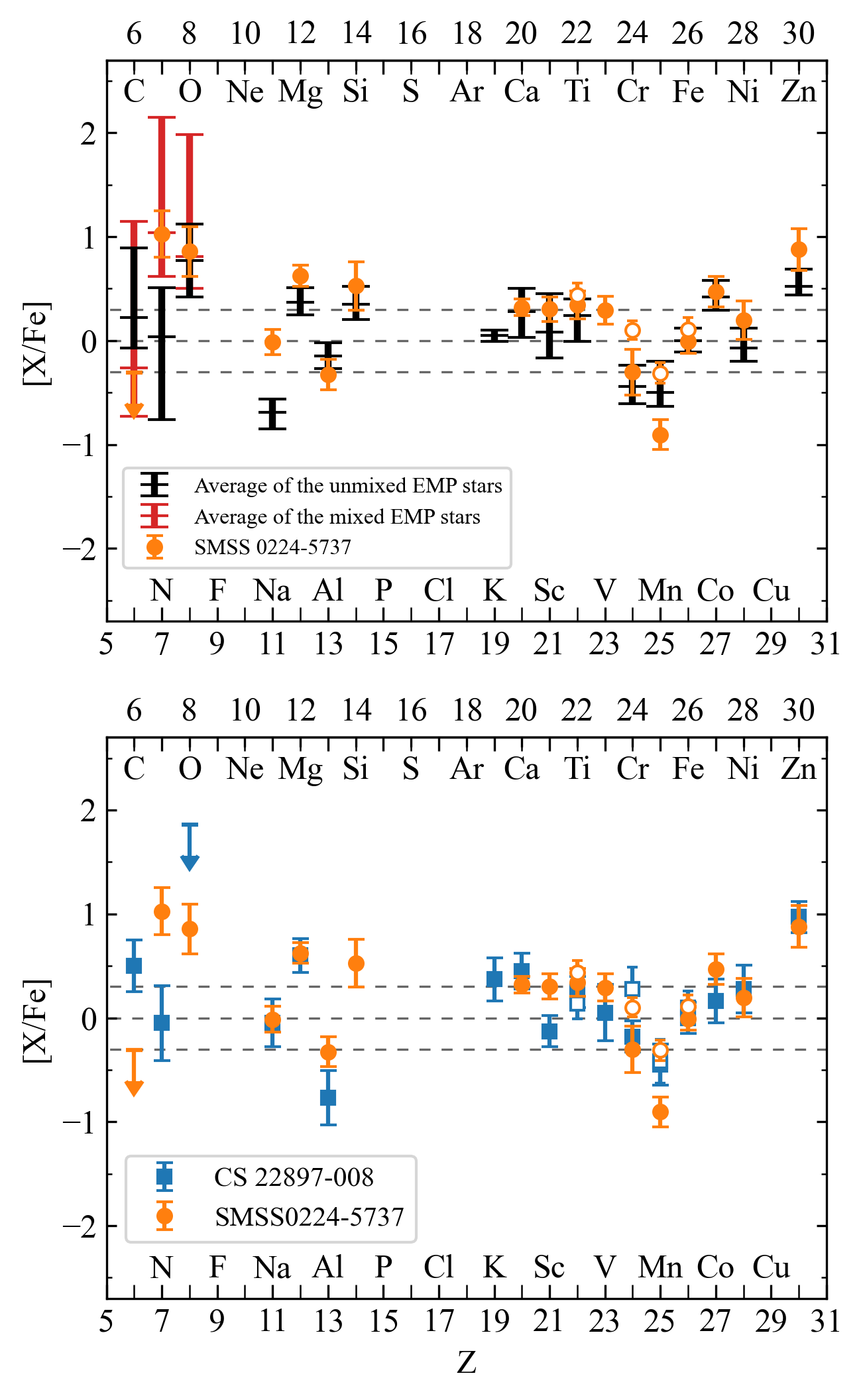}
\caption{Abundance ratios [X/Fe] for light and iron-peak elements as a function of the atomic number (Z) of \starname (orange circles) compared to those of other EMP stars. When abundances are available for both neutral and singly ionized species, the ratios derived from the singly ionized species are plotted as open circles as in Fig.~\ref{fig:abun_all}.
Top panel: Abundances are compared with the averages of unmixed EMP stars (black; \citealt{2004A&A...416.1117C}) and mixed EMP stars (red; \citealt{2005A&A...430..655S,2006A&A...455..291S}). Bottom panel: Comparison with the weak $r$-process star CS~22897-008, which has similar metallicity (blue squares; \citealt{2014AJ....147..136R}). Arrows indicate the upper limits. The horizontal dashed lines represent the solar-scaled abundances ([X/Fe] = 0) and $\pm 0.3$~dex deviations. }
\label{fig:abun_light}
\end{figure}

Figure~\ref{fig:abun_light} shows the abundance patterns of \starname\ up to the Fe-peak elements, that is, the abundance ratios [X/Fe] as a function of the atomic number. The upper panel compares them with the averages of unmixed and mixed EMP stars \citep{2004A&A...416.1117C,2005A&A...430..655S,2006A&A...455..291S}, including a correction for Mg abundance \citep{2009A&A...501..519B}, while the lower panel compares them with CS~22987-008 \citep{2014AJ....147..136R}, one of the EMP stars that shows the largest enhancement of light neutron-capture elements relative to heavy ones (e.g., Zr/Ba), as discussed below. CS~22987–008 exhibits a similar abundance pattern and metallicity ([\ion{Fe}{1}/H] $= -3.83$) to those of \starname.
The abundance pattern of light elements is broadly consistent with those of typical EMP stars.
Among the light elements, the CNO abundance pattern of \starname\ exhibits characteristic features. In contrast to the average of the EMP stars and CS~22987–008, \starname\ exhibits enhancements of N and O in spite of a stringent upper limit on the C abundance. As shown in the top panel of Figure~\ref{fig:abun_light}, this feature is consistent with the abundance pattern of mixed stars, where mixing enhances the conversion of C to N (see also the discussion in Section~\ref{sec:light}). 
Furthermore, in \starname, Na and Mg are slightly enhanced compared with the average of the EMP stars, and a prominent enhancement of Zn is also observed. This feature is similar to that of CS~22987-008 and is typical of an EMP star (bottom panel of Figure~\ref{fig:abun_light}).

\begin{figure}[t]
\centering
\includegraphics[width=\columnwidth]{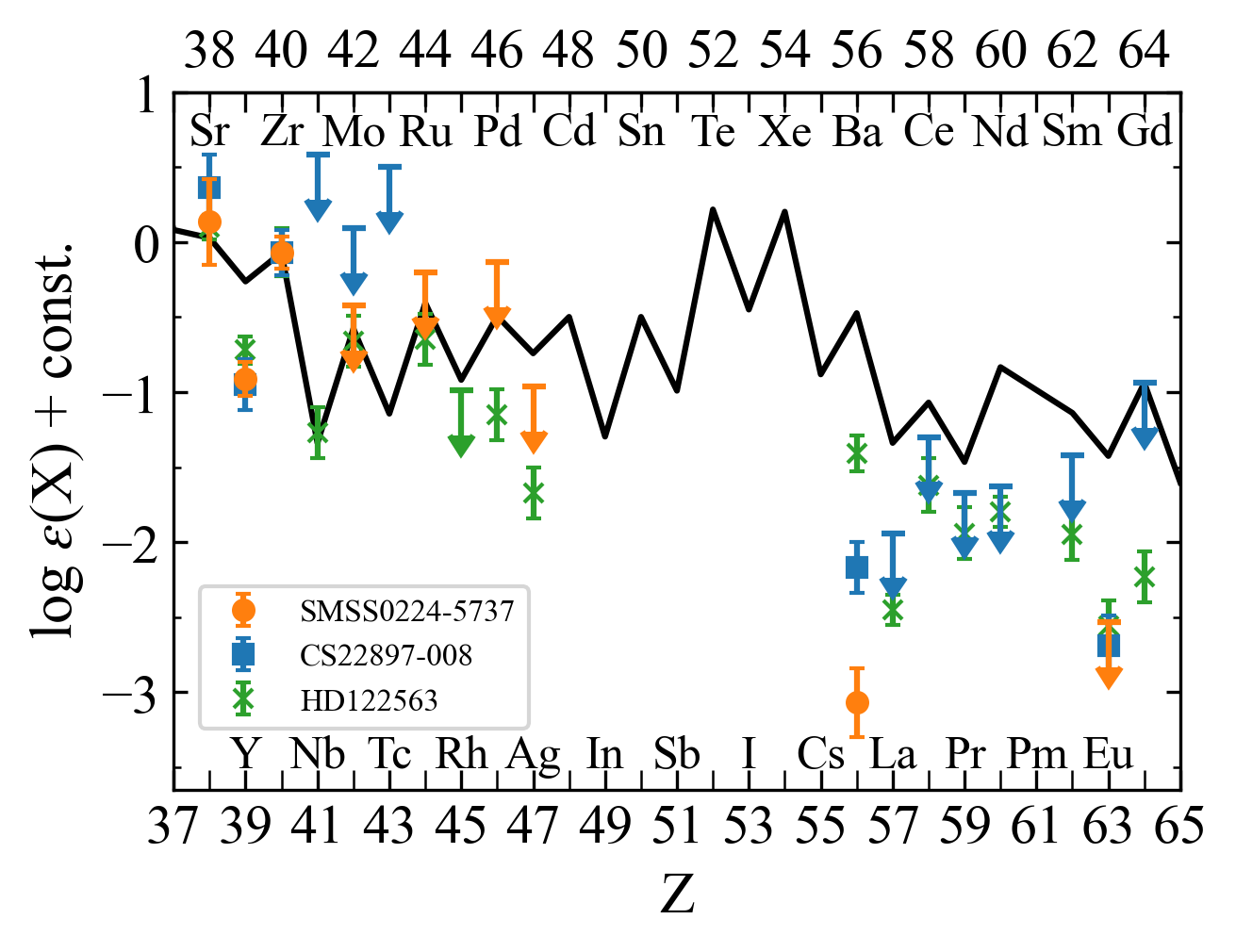}
\caption{Comparison of the abundances of neutron-capture elements among the solar residual $r$-process pattern (black line, \citealt{2004ApJ...617.1091S}), \starname\ (orange symbols), CS~22897-008 (blue symbols, \citealt{2014AJ....147..136R}), and a well-known weak $r$-process star HD~122563 (green symbols, \citealt{2006ApJ...643.1180H}). The solar residual $r$-process pattern and the abundance patterns of HD~122563 and CS~22897-008 are shifted to match the Zr abundance of \starname. The filled circles with error bars represent the detected elements, and the downward arrows indicate the upper limits. The abundance pattern of \starname\ shows enhancements in the first $r$-process peak (e.g., Sr, Y, and Zr) and deficiencies in second peak and lanthanides (e.g., Ba and Eu)}
\label{fig:abun_ncap}
\end{figure}

Figure~\ref{fig:abun_ncap} shows the abundance ratios of neutron-capture elements normalized by the abundance of Zr. The abundance pattern is compared with the solar residual $r$-process pattern  \citep{2004ApJ...617.1091S} and with those of HD~122563 \citep{2006ApJ...643.1180H} and CS~22987-008. 
The most characteristic feature of \starname\ is the abundance of Ba normalized by Zr, which is lower than that of the solar residual $r$-process pattern and those of HD~122563 and CS~22987-008 by $\sim 2.5$~dex, $\sim 1.5$~dex, and $\sim 1$~dex, respectively. 
The abundance ratios among Sr, Y, and Zr in \starname\ are consistent with those in HD~122563 and CS~22987-008. The low Y abundance of \starname, HD~122563, and CS~22987-008 compared to the solar residual $r$-process pattern is likely to stem from the NLTE effect on the \ion{Y}{2} lines \citep{2023MNRAS.525.3718S}.
The low upper limits on Ag and Eu demonstrate low abundances of elements with $Z > 47$, whereas the upper limits on Mo, Ru, and Pd are consistent with the solar residual $r$-process pattern. The low upper limit of the Eu abundance illustrates that the Eu abundance normalized by Zr of \starname\ is lower than or similar to those of HD~122563 and CS~22987-008.

\begin{figure*}[t]
\centering
\includegraphics[width=\textwidth]{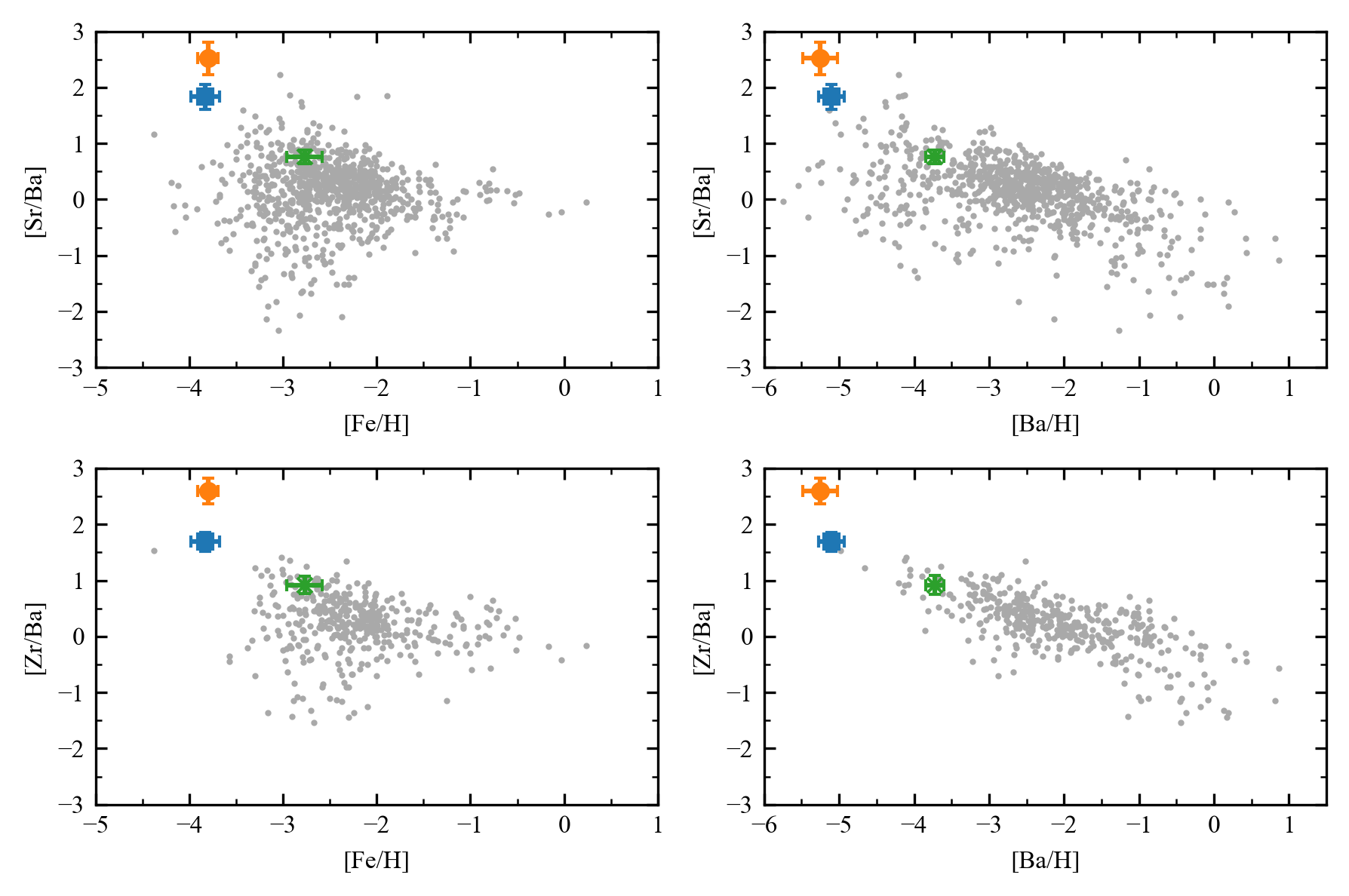}
\caption{Abundance ratios [(Sr, Zr)/Ba] as a function of metallicity [Fe/H] and Ba abundance [Ba/H] for \starname\ (orange), HD~122563 (green), CS~22897-008 (blue), and the stars compiled by the SAGA database (gray). }
\label{fig:abun_SrBa}
\end{figure*}

This results in exceptionally high [Sr/Ba] $= +2.52$ and [Zr/Ba] $= +2.60$ of \starname. Figure~\ref{fig:abun_SrBa} shows the abundance ratios [Sr/Ba] and [Zr/Ba] as functions of metallicity [Fe/H] and Ba abundance [Ba/H] for \starname, together with HD~122563, CS~22897-008, and EMP stars from the SAGA database. Although \starname\ and CS~22987-008 have similar [Fe/H] and [Ba/H] ratios, both are much lower than those of HD~122563, and the [Zr/Ba] ratio of \starname\ is higher than that of CS~22987-008 by $\sim1$~dex. These features make \starname\ an extreme example of the EMP star with weak $r$-process signature. 

\section{Discussion} \label{sec:dis}

\subsection{Where Barium in \texorpdfstring{\starname}{starname} comes?}

We detect Ba in \starname\ for the first time and determine the low abundance of Ba, [Ba/H]~$=-5.25$. In this subsection, we discuss the origin of Ba in \starname. 

One possible explanation for the low Ba abundance in metal-poor stars is the accretion of metals from the interstellar medium (ISM). We assess whether this mechanism could account for the observed Ba abundance in \starname.

The ISM accretion can increase the metal abundance in main-sequence stars to typical values of [X/H] $\sim -6$ to $-5$, with rare cases reaching [X/H] $= -2$ \citep{2017MNRAS.469.4012S}. The surface convective zone mass increases significantly as the star evolves from $\sim10^{-3}$\Msun\ in the main sequence to $\sim10^{-1}$\Msun\ in the red giant branch \citep[e.g.,][]{1995ApJ...444..175F,2004ApJ...611..476S}. The stellar parameters indicate that \starname\ is on the red giant branch. Furthermore, the observed C depletion and low [C/N] ratio are consistent with deep mixing, similar to those observed in red giants (\citealt{2014ApJ...797...21P}, Section~\ref{sec:light}).

Given that typical ISM accretion enriches the [Ba/H] ratio of giants only to $\sim -8$ to $-7$ \citep{2017MNRAS.469.4012S}, which is much lower than the observed [Ba/H]~$= -5.25$, we conclude that the ISM accretion cannot explain the Ba abundance in \starname. 
Instead, Ba must be synthesized during a prior nucleosynthesis event.

\cite{2013AJ....145...26R} pointed out that Ba has been detected in most of metal-poor giants and that all upper limits are consistent with the lowest detected Ba abundance,
so called a Ba abundance floor at [Ba/H]~$\sim -6$. They suggested that a neutron-capture process operates as often as the nucleosynthesis event of elements up to Fe-peak elements in the early Universe, although no nucleosynthesis site is specified. 

The abundance ratio [Ba/H] of \starname\ is six times higher than the lowest detected [Ba/H] among the EMP stars (CS22885-096, \citealt{2014AJ....147..136R}). Therefore, the Ba abundance of \starname\ cannot be explained by a scenario in which Ba is uniformly distributed, that is, the Ba abundance floor, but requires nonuniformly distributed Ba or an additional source of Ba. Furthermore, \starname\ exhibits the high Sr abundance, in contrast to CS22885-096 with [Sr/H]~$=-6.16$ and [Sr/Ba]~$\sim0$. A specific nucleosynthesis event is required to explain the abundances of Sr, Y, and Zr in \starname. Although our measurements cannot exclude the possibility that the non-uniformly distributed Ba abundance floor and the other events synthesizing Sr, Y, and Zr are attributed to the abundance of \starname, there is no reason to prefer multiple origins to a single origin, given its low metallicity.

\subsection{Comparison with Nucleosynthesis Models} \label{sec:ZrBa}

The sharp decrease in the abundances beyond Zr and the low [Ba/Fe] ratio suggest that \starname\ is formed from gas enriched primarily by a nucleosynthesis event in which the main $r$-process is truncated. The large scatter in [Zr/Ba] at low [Ba/H] indicates a variation in $r$-process \citep{2007A&A...476..935F,2013ApJ...766L..13A}. The highest [Zr/Ba] ratio of \starname\ places it as the EMP star with the most pronounced weak $r$-process signatures.
These features provide critical constraints on potential astrophysical sites. We compare the abundance pattern of \starname, particularly the [Zr/Ba] ratio and the upper limits for other heavy elements, with four nucleosynthesis models.

\begin{figure*}[htbp]
\centering
\includegraphics[width=\textwidth]{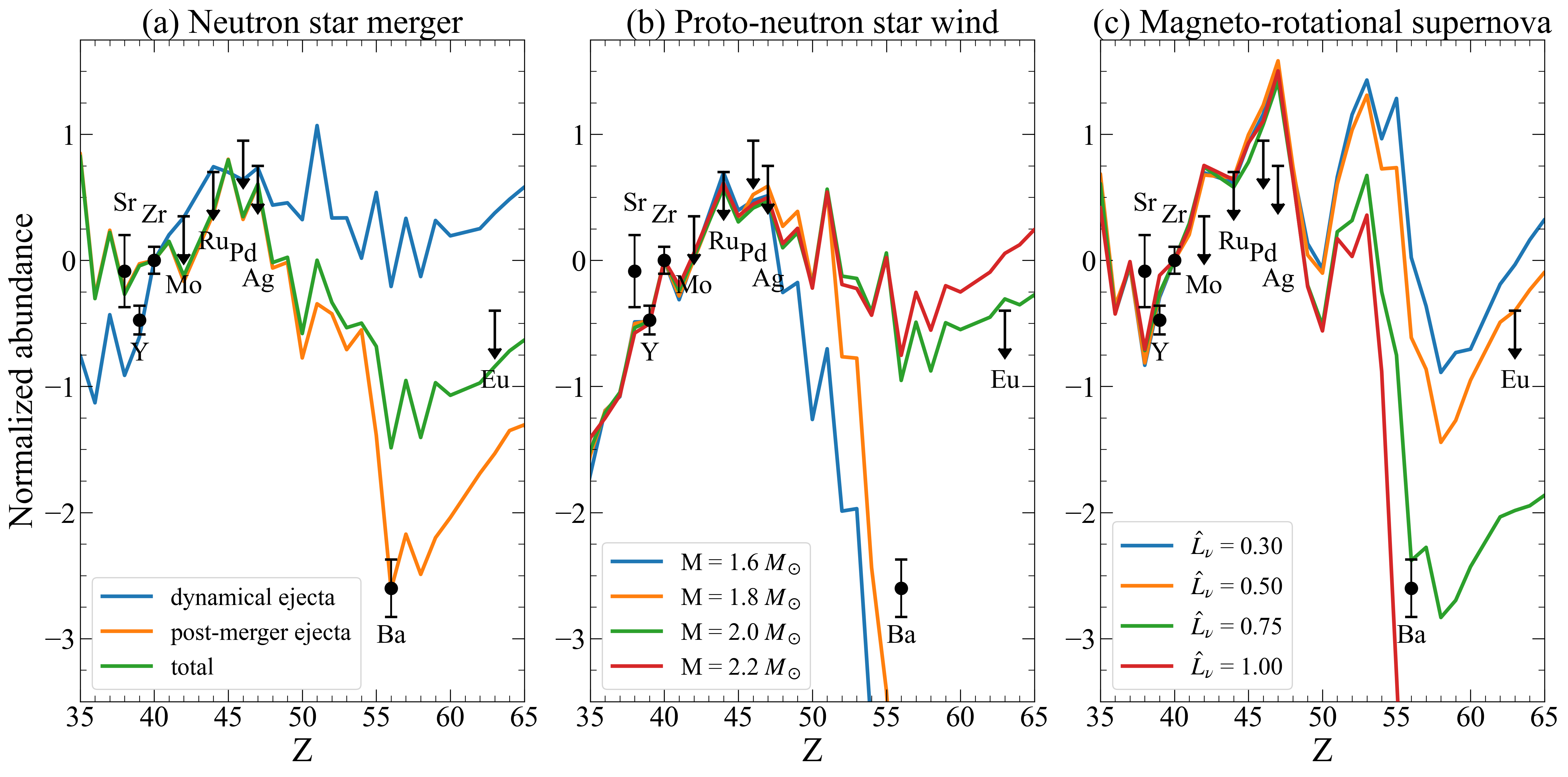}
\caption{Abundance patterns of neutron-capture elements from Sr to Eu of \starname\  compared with the nucleosynthesis models of an NSM involving two 1.35~\Msun\ neutron stars (a, \citealt{2023ApJ...942...39F}), PNS winds with $M_{\rm PNS} = 1.6$--2.4~\Msun\ (b, \citealt{2013ApJ...770L..22W}), and MRSNe with $\Lnu = 0.10$--1.25 (c, \citealt{2017ApJ...836L..21N}). The abundance patterns of the models are vertically shifted to match the [X/Zr] of \starname.}
\label{fig:comp_model}
\end{figure*}

Figure~\ref{fig:comp_model}(a) presents the nucleosynthesis yields of an NSM involving two 1.35~\Msun\ neutron stars, calculated using a general-relativistic simulation \citep{2020ApJ...901..122F}. The NSM model yields dynamical ejecta of $1.5\times10^{-3}$\Msun\ and post-merger ejecta of $7.0\times10^{-2}$\Msun. Whereas the dynamical ejecta in NSMs typically has low \Ye\ and produces heavy $r$-process elements, the post-merger ejecta influenced by strong neutrino irradiation results in high \Ye\ and produces only light $r$-process elements. Figure~\ref{fig:comp_model}(a) illustrates that the [Zr/Ba] ratio of \starname\ is reproduced only if the post-merger ejecta of the NSM model alone contributes to the chemical enrichment of the gas forming \starname. Because the difference in the [Zr/Ba] ratio between the total ejecta of the NSM model and \starname\ is $\sim1$~dex, the abundance ratio of \starname\ might be reproduced only if an NSM yields a negligible amount of dynamical ejecta ($0.2\%$ of the post-merger ejecta) while maintaining the abundance patterns of yields of dynamical and post-merger ejecta. 
Considering the timescale of NSM events, it is difficult for NSMs to occur before the formation of EMP stars (e.g., \citealt{2004A&A...416..997A}). 
Therefore, we conclude that NSMs are unlikely to be the origin of \starname.

Electron-capture supernovae (ECSNe) are also proposed as the weak $r$-process site. Two-dimensional hydrodynamical simulations show that the electron fraction (\Ye) of ECSN ejecta can be as low as 0.4 \citep{2008A&A...485..199J}, and nucleosynthesis under such conditions does not produce heavier elements than Sr \citep{2011ApJ...726L..15W}. The finite Ba abundance observed in \starname\ excludes ECSNe as a viable source unless Ba is synthesized in another event, such as the nonuniformly distributed Ba abundance floor.

Figure~\ref{fig:comp_model}(b) shows the results from semi-analytic, spherically symmetric general-relativistic proto-neutron star (PNS) wind models with $M_{\rm PNS} = 1.6$--$2.2$~\Msun\ \citep{2013ApJ...770L..22W}, assuming a minimum electron fraction of 0.40. The entropy of the wind increases with the PNS mass, and heavier $r$-process elements are synthesized in a higher-entropy environment. The [Zr/Ba] ratio varies largely within the narrow mass range of $M_{\rm PNS} = 1.8$--$2.0$~\Msun, where the observed [Zr/Ba] ratio in \starname\ is reproduced. While most simulations suggest typical PNS masses of $1.2$--$1.6$~\Msun\ \citep{2016ApJ...818..124E}, massive neutron stars up to $\sim 2.0$~\Msun\ have been observed in the Galaxy \citep{2020NatAs...4...72C,2021ApJ...915L..12F}. Thus, \starname\ may have been formed from the ejecta of a core-collapse supernova that produces a massive neutron star.

Figure~\ref{fig:comp_model}(c) shows the results of axisymmetric magneto-hydrodynamical simulations of magneto-rotational supernovae (MRSNe), including neutrino heating \citep{2017ApJ...836L..21N}. The $r$-process nucleosynthesis depends on the neutrino luminosity. Here the fraction of neutrino luminosity $\Lnu$ of a model to a nominal model is adopted as a parameter. The explosion of the nominal model succeeds only with the neutrino heating mechanism. The models with $\Lnu = 0.10$--$1.00$ produce a wide range of [Zr/Ba] ratios. The observed value in \starname\ is well matched by the model with $\Lnu = 0.75$, suggesting that MRSNe under moderate neutrino irradiation can be the origin of \starname. 
The models predict significantly higher Mo, Ru, Pd and Ag abundances than the upper limits obtained in the present study, indicating that the current models do not explain the abundance pattern of \starname. However, the model calculations for the abundances remain uncertain because of uncertainties in the nuclear reaction rates \citep{2012PhRvC..85d8801N}. Improvement of model calculations with more accurate reaction rates is desirable for examining this scenario.

\subsection{Light Elements and Origin of \texorpdfstring{\starname}{starname}} \label{sec:light}

\starname\ exhibits a C-poor and NO-rich nature (Figure~\ref{fig:abun_light}). Although the reduction in C abundance owing to stellar evolution is weak for \Teff\ and \logg\ of \starname\ \citep{2014ApJ...797...21P}, the low [C/N] ratio is consistent with the characteristics of mixed stars \citep{2005A&A...430..655S,2006A&A...455..291S}. Together with the non-detection of Li lines, it is likely that mixing is enhanced in \starname\ compared to EMP stars with the same \Teff\ and \logg, and that the enhanced mixing triggers the conversion of C to N, reduces the C abundance, and enhances the N abundance. If N is converted from C, it is more appropriate to adopt [(C+N)/Fe] instead of [C/Fe] and [N/Fe] to assess the nature of \starname\ at the formation. The upper limit of the [(C+N)/Fe] ratio of \starname\ is $\sim+0.5$. The 3D correction for the OH lines at 3100\AA\ is $\sim-0.2$~dex for \starname\ \citep{2010A&A...519A..46G}. The 3D-corrected [O/Fe] ratio is $\sim+0.66$.

The enhancement of Zn in \starname\ is notable among the EMP stars, although it lies along the trend of EMP stars. The enhancement is attributed to explosive nucleosynthesis in high-entropy environments, possibly due to high-energy core-collapse supernovae such as hypernovae and/or jet-induced explosions \citep[e.g.,][]{2009ApJ...690..526T,2013ARA&A..51..457N}. In contrast, the high [Zn/Fe] ratios are not produced in the NSM models \citep{2023ApJ...942...39F}.

The abundance pattern of the elements from CNO to Fe-peak elements of \starname\ is consistent with those of C-normal EMP stars. Such abundance patterns are well reproduced by high-energy core-collapse supernovae \citep[e.g.,][]{2014ApJ...785...98T}. This is consistent with the conclusion in Section~\ref{sec:ZrBa}, that is, an MRSN can be the origin of \starname, because the MRSN is a mechanism that produces energetic jet-like explosions \citep{2009ApJ...691.1360T}. The low-metallicity nature of \starname\ also favors a scenario in which one nucleosynthesis event simultaneously explains the abundance pattern up to the Fe-peak elements and the high [Zr/Ba] ratios. In conclusion, we propose that the origin of \starname\ is likely an MRSN.

\section{Summary} \label{sec:sum}

We analyze the high-resolution spectra obtained using VLT/UVES and determine the elemental abundances of the EMP star with a weak $r$-process signature, \starname. We measure the abundances of 26 elements, of which five (N, O, V, Zn, and Ba) are detected for the first time. In addition, we derive upper limits for Mo, Ru, Pd, Ag, and Eu.

In contrast to the previous work \citep{2015ApJ...807..171J}, we find no CH molecular lines and place an upper limit on the C abundance ([C/Fe] $< -0.31$), whereas the enhancements of N and O are found. Although the enhancement of N and the reduction of C owing to stellar evolution are weak for stars with \Teff\ and \logg\ similar to \starname\ \citep{2014ApJ...797...21P}, the characteristics of \starname\ are consistent with those of mixed stars \citep{2005A&A...430..655S,2006A&A...455..291S}. This indicates that the mixing during the stellar evolution alters the surface abundance of \starname.

The $\alpha$ and Fe-peak elements generally follow the trends of EMP stars, and the abundance pattern up to the Fe-peak elements is roughly consistent with the average of mixed EMP stars. A notable feature is the high [Zn/Fe] ratio ([Zn/Fe] $= +0.88$), which also follows the trend for EMP stars.

We confirm the enhancement of the first-peak neutron-capture elements (Sr, Y, and Zr) relative to Fe. The abundance ratios among Sr, Y, and Zr are similar to those of HD~122563 and CS~22987-008, which show the weak $r$-process signatures. The most remarkable feature of \starname\ is that Ba is extremely deficient but present. We first determine the Ba abundance [Ba/H] $= -5.25$, resulting in [Ba/Fe] $= -1.45$. No lines of neutron-capture elements heavier than Zr are detected, except for Ba, and we put a stringent upper limit of [Eu/H] $< -3.05$. As a result, the extremely high ratios [Sr/Ba] $= +2.52$ and [Zr/Ba] $= +2.60$ place \starname\ the EMP star with the most pronounced weak $r$-process signatures observed to date.

The extremely high [Zr/Ba] ratio and other abundance ratios provide clues to the origin of \starname. We compare the abundance pattern of neutron-capture elements with the yields of the nucleosynthesis models. The sharp decline in abundances beyond Zr and the very low Ba abundance disfavor the NSM or ECSN models. We note that the ECSN model can explain the abundance of \starname\ only if the non-uniformly distributed Ba abundance floor is created by another nucleosynthesis event. 
Instead, the observed abundance pattern of neutron-capture elements can be reproduced either by PNS wind models in a narrow mass range or by MRSN models with moderate neutrino luminosity. Because the PNS wind model requires fine tuning, the high [Zn/Fe] ratio is consistent with the MRSN model, and the low-metallicity nature of \starname\ favors a single origin over multiple origins, we conclude that the MRSN is the most plausible origin of \starname.

This study demonstrates that the abundance measurements of both light and neutron-capture elements, even with low abundances, are crucial for unveiling the astrophysical sites of the weak $r$-process. Future systematic deep spectroscopic studies of EMP stars exhibiting the weak $r$-process signatures are in high demand.


\begin{acknowledgments}
We thank T. Matsuno and N. Christlieb for constructive comments on the abundance analysis. We also acknowledge S. Fujibayashi, N. Nishimura, and S. Wanajo for providing theoretical yields and valuable discussions. 
This work was supported by JSPS KAKENHI Grant Numbers 21H04499, 23H04894, 24K00682, 25K01046, 25H02196, and 25H00674.
H.O. was supported by the NAOJ Overseas Visit Program for Young Researchers (FY2024) and a grant from the Hayakawa Satio Fund awarded by the Astronomical Society of Japan.
This research was based on observations collected at the European Southern Observatory under ESO programme 0101.D-0677(A).
This research was made possible through the use of the AAVSO Photometric All-Sky Survey (APASS), funded by the Robert Martin Ayers Sciences Fund and NSF AST-1412587.
This work has made use of data from the European Space Agency (ESA) mission {\it Gaia} (\url{https://www.cosmos.esa.int/gaia}), processed by the {\it Gaia} Data Processing and Analysis Consortium (DPAC, \url{https://www.cosmos.esa.int/web/gaia/dpac/consortium}). Funding for the DPAC has been provided by national institutions, in particular the institutions participating in the {\it Gaia} Multilateral Agreement.
\end{acknowledgments}

\bibliography{sample631}{}
\bibliographystyle{aasjournal}



\end{document}